\documentclass[reprint,aps,prl,secnumarabic,superscriptaddress]{revtex4-2}
\usepackage{graphicx}
\usepackage{amsmath}
\usepackage{bm}
\usepackage[percent]{overpic}
\usepackage{float}
\usepackage{caption}
\usepackage{ragged2e}
\usepackage{amssymb}
\usepackage{subcaption}
\usepackage[colorlinks=false]{hyperref}

\begin{document}

\title{Adiabaticity Crossover: From Anderson Localization to Planckian Diffusion}

\author{Tiange~Xiang}
\affiliation{School of Physics, Peking University, Beijing 100871, China}

\author{Yubo~Zhang}
\affiliation{Quantum Science and Engineering, Harvard University, Cambridge, Massachusetts 02138, USA}

\author{Joonas~Keski-Rahkonen}
\affiliation{Department of Chemistry and Chemical Biology, Harvard University, Cambridge, Massachusetts 02138, USA}
\affiliation{Department of Physics, Harvard University, Cambridge, Massachusetts 02138, USA}

\author{Anton~M.~Graf}
\affiliation{Harvard John A. Paulson School of Engineering and Applied Sciences, Harvard University, Cambridge, Massachusetts 02138, USA}
\affiliation{Department of Chemistry and Chemical Biology, Harvard University, Cambridge, Massachusetts 02138, USA}

\author{Eric~J.~Heller}
\affiliation{Department of Chemistry and Chemical Biology, Harvard University, Cambridge, Massachusetts 02138, USA}
\affiliation{Department of Physics, Harvard University, Cambridge, Massachusetts 02138, USA}

\date{\today}

\begin{abstract}

We investigate electron transport in one dimension from the quantum-acoustic perspective, where the coherent-state representation of lattice vibrations results in a time-dependent deformation potential whose rate is set by the sound speed, fluctuation spectrum is set by the temperature, and overall amplitude is set by the electron-lattice coupling strength. We introduce an acceleration-based adiabatic criterion, consistent with the adiabatic theorem and Landau--Zener theory, that separates adiabatic and diabatic dynamics across the $(T,v)$ plane. The discrete classification agrees with a continuous mean-squared acceleration scale and correlates with a coherence measure given by the ratio of coherence length to the initial packet width $L_\phi(t)/\sigma_0$. We identify a broad Planckian domain in which the dimensionless diffusivity $\alpha\!=\!Dm/\hbar$ is of order unity and only weakly depends on the parameters. This domain is more prevalent in diabatic regions and in areas of reduced phase coherence, indicating a dephasing driven crossover from Anderson localization to Planckian diffusion. Using the Einstein relation together with nearly constant $\alpha$, we directly obtain a low temperature tendency $1/\tau_{\rm tr}\propto T$, offering a insight to $T$-linear resistivity in strange metals. These results provide a unified picture that links adiabaticity, dephasing, and Planckian diffusion in dynamically disordered quantum-acoustics.

\end{abstract}

\maketitle

\setcounter{secnumdepth}{2}
\renewcommand\thesection{\Roman{section}}
\renewcommand\thesubsection{\Alph{subsection}}

\section{Introduction}

Anderson localization is a paradigmatic consequence of phase coherent interference in disordered media. In one dimension any static disorder localizes single particle eigenstates and suppresses transport \cite{PhysRev.109.1492,RevModPhys.57.287,PhysRevLett.42.673,J_T_Edwards_1972}. The phenomenon is broadly observed across platforms including photonic lattices and ultracold atoms, which underscores its generality \cite{Schwartz2007,Billy2008,doi:10.1126/science.1209019,White2020}. When the disorder evolves in time the interference corrections that sustain localization are suppressed and diffusion can reemerge \cite{10.1063/1.5054017}. A central open question is how the loss of phase coherence organizes real time dynamics in driven landscapes and how this connects to transport \cite{RevModPhys.80.1355}.

The adiabatic theorem and the Landau--Zener framework provide guidance for dynamics under slow parameter changes \cite{Born1928,doi:10.1143/JPSJ.5.435,10.1098/rspa.1932.0165,Landau1932_PZS_LZ,Stueckelberg1932_HPA,Majorana1932,Shevchenko2010_PhysRep_LZSReview}. Adiabatic passage across an avoided crossing follows the adiabatic eigenstate and therefore switches diabatic channels, whereas diabatic passage follows the original channel. In crystalline solids time dependence naturally arises from lattice vibrations. From the quantum-acoustic perspective, these vibrations act as a deformation potential whose rate of change in time is set by the sound speed, fluctuation spectrum is set by the temperature through thermal mode populations, and overall amplitude is set by the electron-lattice coupling strength \cite{PhysRevB.106.054311}. Recent work has used this coherent-state picture to analyze Planckian resistivity and the role of lattice-driven dephasing in metals and model systems \cite{doi:10.1073/pnas.2404853121,e26070552}.

Standard diagnostics of adiabaticity typically evaluate instantaneous gaps and nonadiabatic couplings or invoke two-level Landau--Zener probabilities \cite{10.1063/1.2798382}. In chemical physics, mixed quantum-classical surface hopping triggers stochastic hops between adiabatic surfaces based on coupling strength and fewest-switches logic \cite{10.1063/1.459170}. For a multi-mode, time-dependent deformation potential, avoided crossings are dense and no single sweep rate characterizes the drive. Repeated diagonalization becomes costly and basis dependent, and classical surface assumptions obscure phase information. This motivates basis-independent observables that diagnose adiabatic or diabatic motion directly in real time.

Despite extensive work on dephasing and on adiabatic transitions, a quantitative bridge that turns adiabaticity into an practical indicator of phase coherence and ties it to transport in dynamically disordered landscapes remains underdeveloped. In particular there is no single framework that links adiabatic versus diabatic motion, erosion of phase coherence by fluctuations, and the emergence of Planckian diffusion \cite{RevModPhys.94.041002}.

Here we develop such a framework in one dimension. We drive an electronic wave packet with a time-dependent deformation potential and introduce an acceleration-based adiabatic criterion that is consistent with the adiabatic theorem and Landau--Zener theory. We plot regime maps across temperature and sound speed, compare this discrete classification with a continuous measure from the mean-squared acceleration, and quantify residual phase coherence through the dimensionless ratio of coherence length to the initial packet width. We then connect coherence loss to transport and identify a broad Planckian domain in which the dimensionless diffusivity $\alpha\!=\!Dm/\hbar$ is of order unity \cite{zhang2024planckiandiffusionghostanderson}. The picture that emerges is that adiabaticity tracks how much phase coherence the system retains, while diabatic dynamics correlate with dephasing that drives a crossover from Anderson localization to Planckian diffusion. We outline implications for low temperature strange metal phenomenology and motivate experiments that correlate phase-coherence measures with transport in low-dimensional materials subject to engineered dynamic disorder \cite{doi:10.1126/science.1227612,Legros2019,Grissonnanche2021}.

\section{Methods}

\subsection{Quantum-Acoustic Deformation Potential and Electronic Dynamics}

We investigate electron transport in one dimension from the quantum-acoustic perspective, where a coherent-state representation of lattice vibrations results in a time-dependent deformation potential that acts on a single electronic wave packet \cite{PhysRevB.106.054311}. When there is no external electromagnetic field in the system, the electronic state obeys
\begin{equation}
    i\hbar\,\partial_t \psi(x,t)=\left[-\frac{\hbar^2}{2m}\,\partial_x^2+V_D(x,t)\right]\psi(x,t)
\end{equation}
The deformation potential is synthesized on a discrete $q$ grid from coherent-state amplitudes $\{\alpha_q(t)\}$,
\begin{equation}
    V_D(x,t)=\sum_{q\neq 0} 2\,\mathrm{Re}\!\left[g_q\,\alpha_q(t)\,e^{iqx}\right]
\end{equation}
with $\alpha_q(0)$ drawn from a thermal coherent ensemble with random phases. Within the random phase approximation (RPA), the initial coherent amplitudes are sampled with independent random phases
\begin{equation}
    \alpha_q(0)=\sqrt{N_q}\,e^{i\varphi_q},\qquad
    \varphi_q \sim \mathcal{U}(0,2\pi)
\end{equation}
and mode occupations obey a Bose--Einstein distribution
\begin{equation}
    N_q=\frac{1}{e^{\hbar\omega_q/k_BT} - 1}
\end{equation}
where the overall normalization is absorbed into $g_q$, so the overall amplitude of $V_D(x,t)$ is controlled solely by the electron-lattice coupling strength $g_q$. The coherent amplitudes evolve with full back-action according to
\begin{equation}
    i\hbar\,\frac{\mathrm{d}\alpha_q(t)}{\mathrm{d}t}=\hbar\omega_q\,\alpha_q(t)+g_q\,n_q(t)
\end{equation}
which is the continuous-time form implemented in the code as the update
\begin{equation}
    \alpha_q(t+\mathrm{d}t)=\alpha_q(t)\,e^{-i\omega_q\mathrm{d}t}-i\,\frac{\mathrm{d}t}{\hbar}\,g_q\,n_q(t)
\end{equation}
Here $n_q(t)$ denotes the $q$-component of the instantaneous electron density. In our adiabaticity study we generally disable back-action so that the coherent state evolve simply as $\alpha_q(t)=\alpha_q(0)\,e^{-i\omega_q t}$. This isolates temporal dephasing generated by the lattice field itself and corresponds to the pre-self-trapping stage of the dynamics before a polaron forms \cite{doi:10.1073/pnas.2426518122}. A polaron implies that the wave packet becomes trapped in a self-consistent potential well whose envelope in our numerics is well approximated by a Bessel-like profile. The electron is propagated by a split-operator FFT scheme on a periodic spatial grid. Specifically, we adopt an acoustic dispersion
\begin{equation}
    \omega_q=\frac{2v}{a}\,\Big|\sin\!\frac{q a}{2}\Big|
\end{equation}
where $v$ is the sound speed and $a$ is the lattice constant. The electron-lattice coupling strength $g_q$ is parameterized by a single energy scale $g$ and the mode-resolved coupling used in the code is
\begin{equation}
    g_q = g\,\frac{s}{\sqrt{(L/\pi)\,\big|\sin(\pi s/L)\big|}},\qquad
    q=\frac{2\pi s}{L a},\qquad |q| < q_D
\end{equation}
with $L$ the number of unit cells in the supercell and $s\in\mathbb{Z}$ the integer mode index with both signs retained within the cutoff. The mode set is truncated at the Debye wave vector $q_D$, which we choose with reference to the Fermi wave vector $k_F$ so that the ultraviolet content of $V_D$ is selected in accordance with the underlying electronic scale.

In summary, $v$ fixes the rate of change in time through $\omega_q$, $T$ fixes the fluctuation spectrum through thermal occupations of $\alpha_q$, and $g$ fixes the overall amplitude scale of $V_D(x,t)$ via $g_q$.

\subsection{Acceleration-Based ``Kink'' Criterion of Adiabaticity}
Across temperature $T$, sound speed $v$, and electron-lattice coupling strength $g$, we distinguish adiabatic and diabatic crossovers using an acceleration-based adiabatic criterion that is consistent with the adiabatic theorem and Landau--Zener theory \cite{Born1928,doi:10.1143/JPSJ.5.435,10.1098/rspa.1932.0165,Shevchenko2010_PhysRep_LZSReview}. Two common diagnostics in the literature are as follows. First, adiabatic-basis tests evaluate instantaneous gaps and nonadiabatic couplings and judge motion adiabatic when $\lvert\langle m|\dot n\rangle\rvert$ is small relative to the gap scale or when the two-level Landau--Zener probability is exponentially small \cite{10.1063/1.2798382}. Second, mixed quantum-classical surface hopping triggers stochastic hops between adiabatic surfaces using the nonadiabatic coupling strength and fewest-switches logic \cite{10.1063/1.459170}. In our system the deformation field is multi-mode and time fluctuating, avoided crossings are dense, there is no single sweep rate, and repeated diagonalization is costly and basis dependent, so these diagnostics become fragile. Surface hopping also presumes well separated potential energy surfaces (PESs) and classical nuclei, whereas we propagate a single quantum wave packet in a continuum-like field and must retain phase information. This motivates a basis-independent, directly observable diagnostic built from the center-of-mass acceleration.

We define the wave packet center and its acceleration as
\begin{equation}
\begin{aligned}
x_{\rm cm}(t) &= \int \mathrm{d}x\, x\, |\psi(x,t)|^{2}\\
a_{\rm cm}(t) &= \frac{\mathrm{d}^{2}x_{\rm cm}(t)}{\mathrm{d}t^{2}}
\end{aligned}
\end{equation}

Regarding the details of implementation, we smooth $|a_{\rm cm}(t)|$ with a short-window median and locate local maxima. For each candidate we estimate a robust local background from the lower half of samples in a neighborhood and form z-scores. A peak is accepted only if four scale-free tests are met: (i) height z exceeds a threshold;
(ii) prominence z above adjacent bases exceeds a threshold; (iii) the width around half-prominence is below a cap; (iv) and the separation from the last accepted peak exceeds a minimum. Thresholds are relative and window sizes scale with the record length, so uniform rescaling of forces or time sampling does not change labels. If at least three peaks are accepted within the observation window the trajectory is labeled adiabatic and the time of the first accepted peak defines $\tau$; otherwise it is labeled diabatic. Physically, adiabatic passage through avoided crossings produces discrete channel switches in a diabatic basis and sharp changes in the expectation value of the force that appear as peaks in $a_{\rm cm}(t)$ and as kinks in $|\psi(x,t)|^{2}$; in strongly diabatic conditions with stochastic temporal fluctuations, random phase accumulation averages the force signal into a smooth trace and reduces phase coherence.

As a scalar activity measure we also use the mean-squared acceleration ($\mathrm{MSA}$)
\begin{equation}
    \mathrm{MSA}=\sqrt{\frac{1}{t_{\mathrm{tot}}}\int_{0}^{t_{\mathrm{tot}}} a_{\rm cm}^{2}(t)\,\mathrm{d}t}
\end{equation}
which correlates with kink density and prominence in parameter scans.

\subsection{Quantifying Phase Coherence: Coherence Length}
To quantify coherence in real space we define the normalized first-order spatial coherence
\begin{equation}
  g^{(1)}(x,x^\prime;t)=\frac{\psi^*(x,t)\,\psi(x^\prime,t)}{\sqrt{|\psi(x,t)|^2\,|\psi(x^\prime,t)|^2}}
\end{equation}
which is the standard field-field (first-order) coherence function in wave physics and quantum optics \cite{PhysRev.131.2766}. and the spatially averaged coherence profile
\begin{equation}
  \mathcal{C}(\Delta x,t)=\Big| \big\langle g^{(1)}(x,x+\Delta x;t)\big\rangle_x \Big|
\end{equation}
a common practice when assessing phase coherence of fluctuating or disordered fields \cite{RevModPhys.71.313,RevModPhys.80.885}. The coherence length is defined as a separation-weighted average
\begin{equation}
  L_\phi(t)=\frac{\int_{0}^{\infty}\! \mathrm{d}(\Delta x)\,\mathcal{C}(\Delta x,t)\,\Delta x}{\int_{0}^{\infty}\! \mathrm{d}(\Delta x)\,\mathcal{C}(\Delta x,t)}
\end{equation}
which is consistent with extracting coherence scales from the decay of the first-order correlation function in experiments ranging from ultracold gases to mesoscopic wave transport \cite{Hadzibabic2006,RevModPhys.80.885,RevModPhys.71.313}.
To obtain a dimensionless and comparable measure of the remaining phase coherence we use the ratio $L_\phi(t)/\sigma_0$, where $\sigma_0$ denotes the spatial standard deviation of the initial electronic Gaussian wave packet. A larger value indicates that phase coherence extends across a span that exceeds the initial packet width, whereas a smaller value signals stronger dephasing relative to the initial state. In the present context $L_\phi$ serves as a coherence measure that tracks the erosion of phase coherence by temporal disorder. While $L_\phi(t)/\sigma_0$ is time dependent, in our simulations it relaxes toward a plateau. We therefore report $L_\phi/\sigma_0$ as the average of $L_\phi(t)/\sigma_0$ over the final 30\% of the simulation window. This procedure removes early transients and yields a time-independent measure that reflects properties of the system rather than the run length. All $L_\phi/\sigma_0$ values quoted below follow this averaging rule. The map of $L_\phi/\sigma_0$ across $(T,v,g)$ aligns with the kink based regime boundary and shows that decreasing adiabaticity correlates with faster loss of phase coherence. This supports the interpretation that increasingly diabatic driving conditions promote the Planckian diffusion regime where localization is suppressed.

\onecolumngrid
\begin{center}
\setlength{\tabcolsep}{4pt}
\begin{tabular}{ccc}
  \begin{overpic}[width=0.32\textwidth]{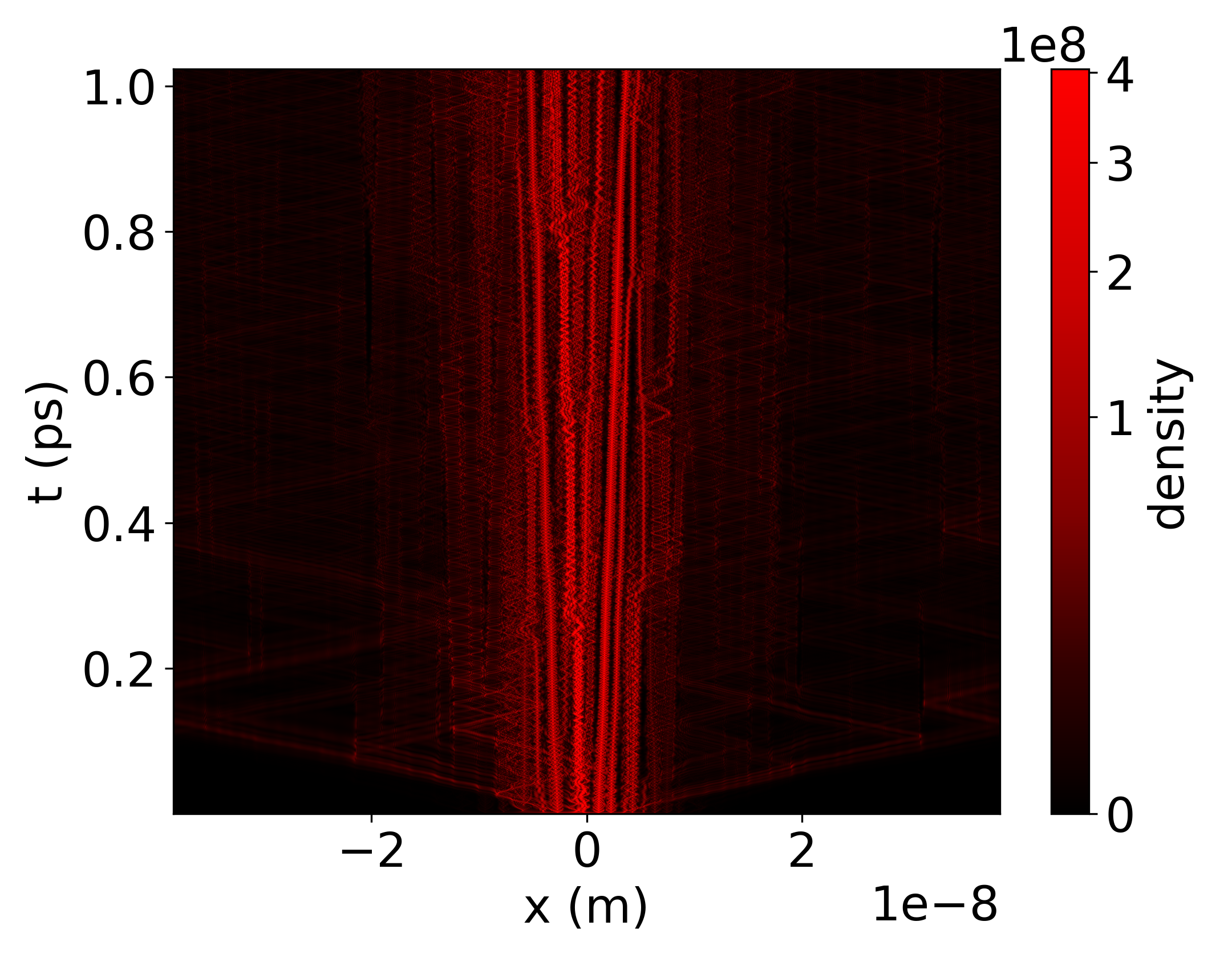}\put(0,76){\small (a)}\end{overpic} &
  \begin{overpic}[width=0.32\textwidth]{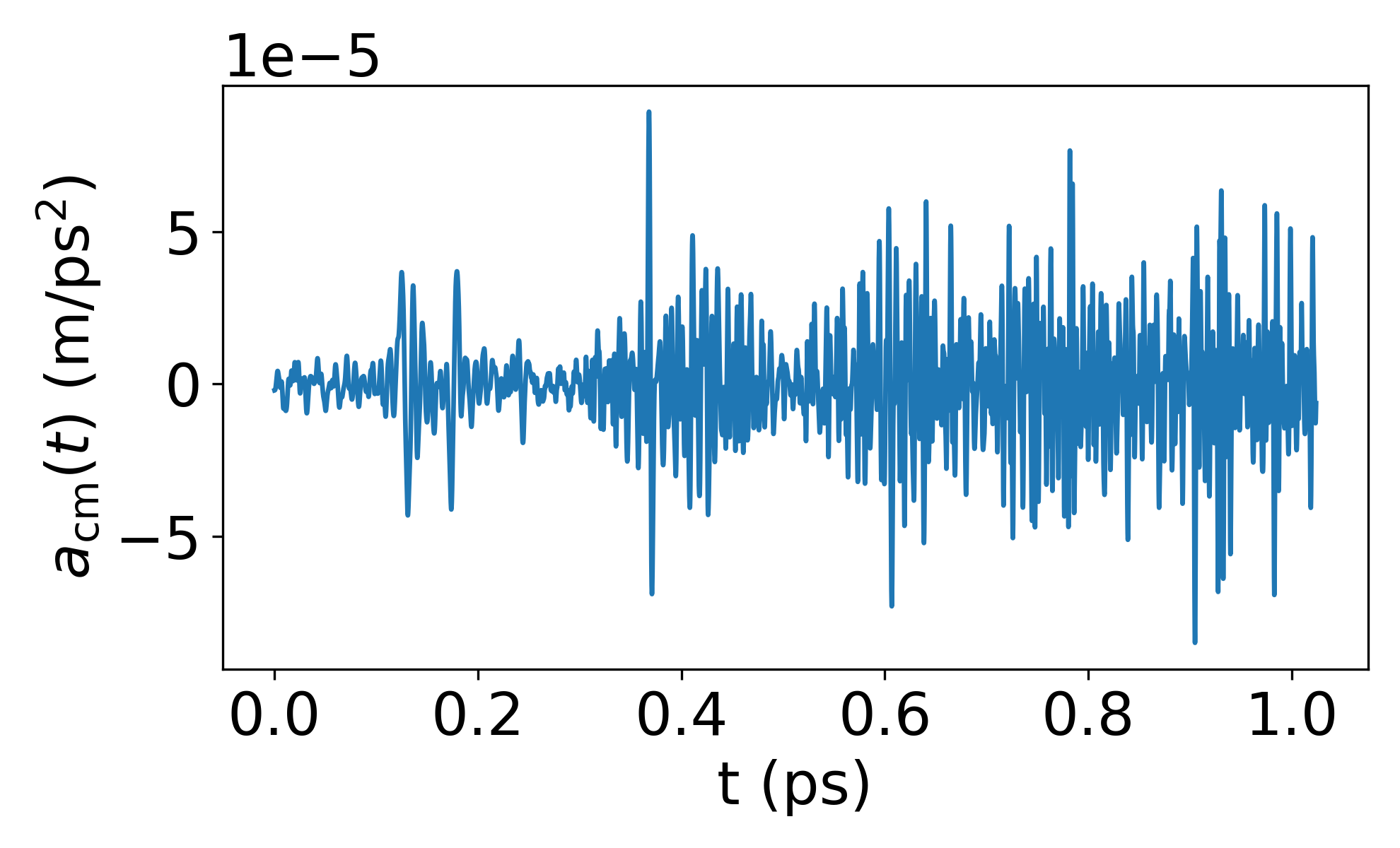}\put(0,76){\small (b)}\end{overpic} &
  \begin{overpic}[width=0.32\textwidth]{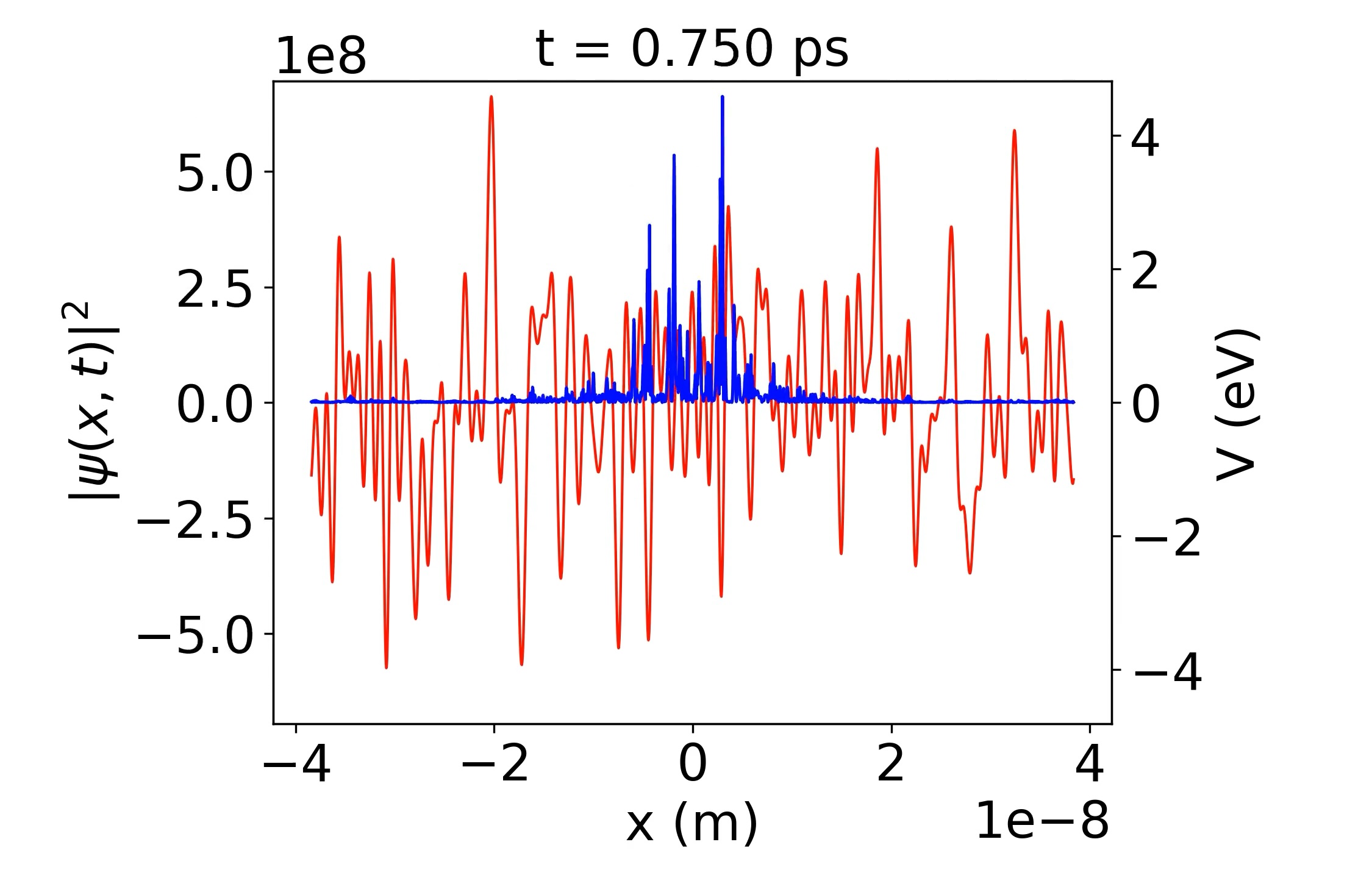}\put(0,76){\small (c)}\end{overpic} \\
  \begin{overpic}[width=0.32\textwidth]{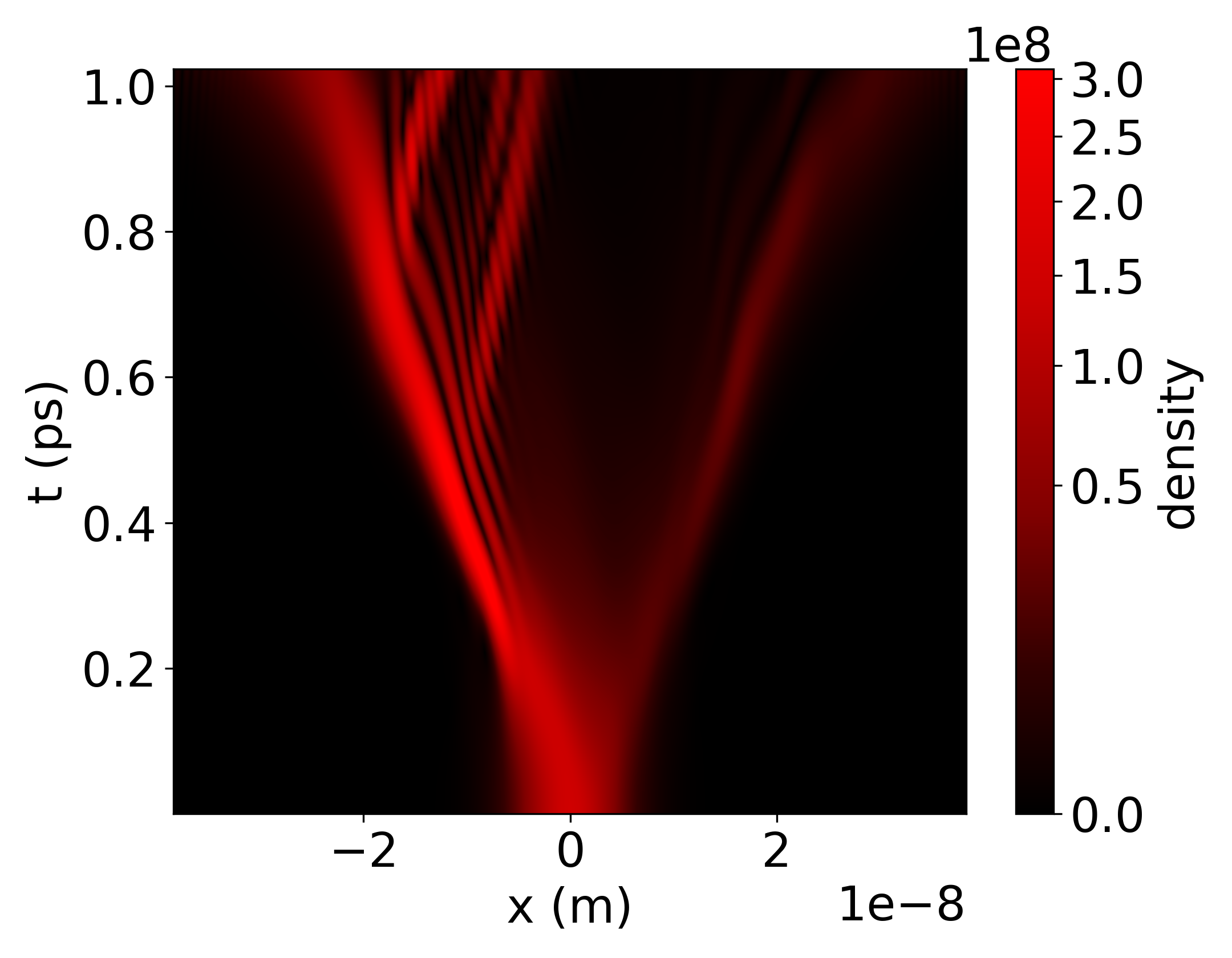}\put(0,76){\small (d)}\end{overpic} &
  \begin{overpic}[width=0.32\textwidth]{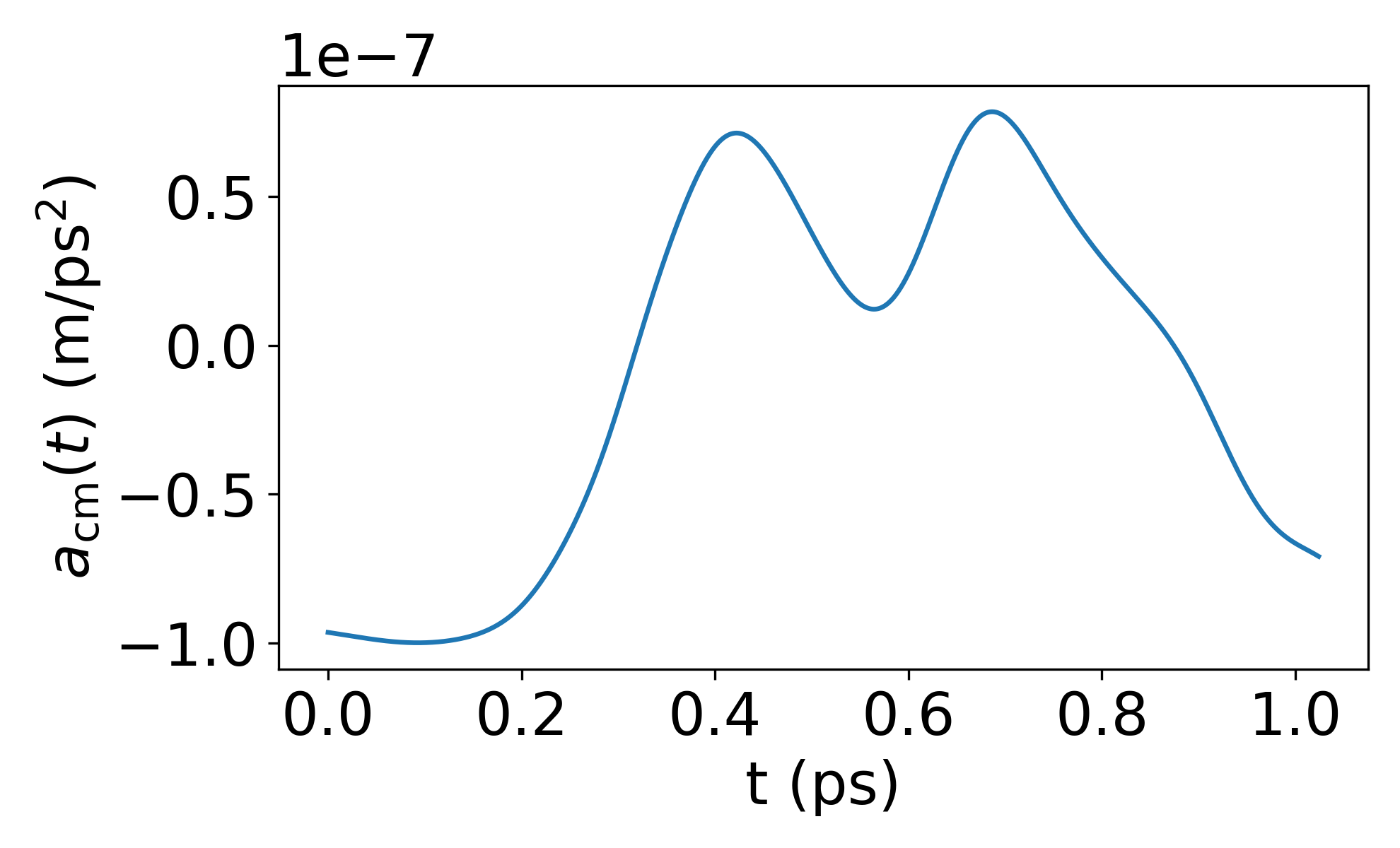}\put(0,76){\small (e)}\end{overpic} &
  \begin{overpic}[width=0.32\textwidth]{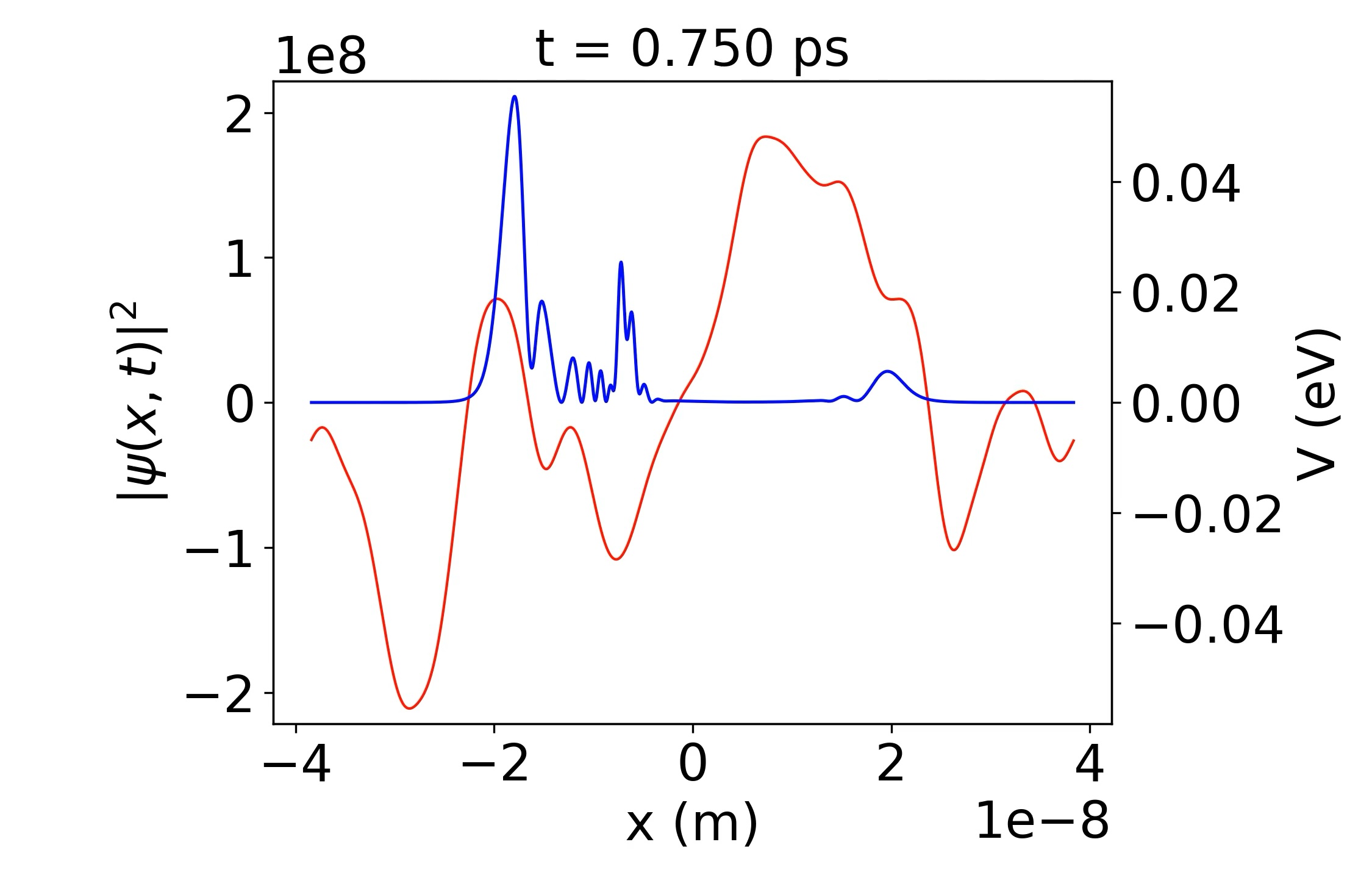}\put(0,76){\small (f)}\end{overpic}
\end{tabular}
\captionsetup{width=\textwidth, justification=RaggedRight, singlelinecheck=false}
\captionof{figure}{\label{fig:fig1}
Acceleration-based adiabatic criterion illustrated with space-time density, center-of-mass acceleration, and equal-time snapshots. The top row shows an adiabatic case and the bottom row shows a diabatic case. Panels (c) and (f) are snapshots at $t=0.750$\,ps where blue curves plot $|\psi(x,t)|^{2}$ and red curves plot the instantaneous deformation potential $V(x,t)$. Panels (a)--(c) share adiabatic parameters and panels (d)--(f) share another set of diabatic parameters.
(a) Adiabatic $x$--$t$ map $|\psi(x,t)|^{2}$ with discrete kinks that mark resolved channel switches between localized pockets. Red indicates the magnitude of $\vert \psi\vert^2$.
(b) Adiabatic $a_{\rm cm}(t)$ with many sharp accepted peaks that pass our scale-invariant peak tests.
(c) Adiabatic snapshot at $t=0.750$\,ps showing a compact probability profile trapped near a pocket of $V(x,t)$.
(d) Diabatic $x$--$t$ map that appears smooth under our thresholds with no resolvable kinks.
(e) Diabatic $a_{\rm cm}(t)$ dominated by broad undulations that fail the peak tests.
(f) Diabatic snapshot at $t=0.750$\,ps showing a more extended and asymmetric probability profile relative to $V(x,t)$.}
\end{center}
\twocolumngrid

\section{Results}

\subsection{Acceleration-Based Adiabatic Criterion: Adiabatic and Diabatic Dynamics}

From the quantum-acoustic perspective, an electronic wave packet evolves in a time-dependent deformation potential \cite{PhysRevB.106.054311,doi:10.1073/pnas.2426518122}. The fluctuation spectrum is set by the Bose--Einstein distribution at temperature $T$, while the rate of change in time follows the acoustic dispersion with the sound speed $v$. The overall coupling to the lattice field is controlled by a single deformation parameter $g$ that descends from the deformation potential constant $E_d$. For fixed $g$ and across the $(T,v)$ plane the dynamics separates into two qualitatively distinct behaviors.

Representative space-time maps and acceleration traces in Figure~\ref{fig:fig1} summarize the two dynamical sectors. In the adiabatic regime [Figure~\ref{fig:fig1} (a--c)] the $x$--$t$ map shows a discrete sequence of kinks that form a clear zigzag track and the center-of-mass acceleration contains many sharp accepted peaks. The equal-time snapshot in Figure~\ref{fig:fig1} (c) at $t=0.750$\,ps reveals a compact probability profile aligned with a pocket of the deformation potential, consistent with channel switches between instantaneous localized eigenstates. In the diabatic regime [Figure~\ref{fig:fig1} (d--f)] the $x$--$t$ map is smooth because the landscape moves too rapidly for projection onto a single instantaneous eigenstate, so no peaks pass our thresholds; the snapshot in Figure~\ref{fig:fig1} (f) at $t=0.750$\,ps shows a broader and more asymmetric profile. Weak incoherent fluctuations then act as an effective dephasing bath that suppresses phase-coherent backscattering and reduces the residual coherence \cite{RevModPhys.57.287,10.1063/1.5054017}. Empirically the diabatic sector occupies a broader portion of the $(T,v)$ map and it is not uniform. Within it lies the Planckian domain where the dimensionless diffusivity $\alpha\!=\!Dm/\hbar$ is of order unity \cite{zhang2024planckiandiffusionghostanderson}, and there are also zones toward larger sound speed $v$ that exhibit even larger diffusivity, consistent with a faster sweep of the deformation potential and stronger dephasing. All of these trajectories fail the peak tests and produce smooth acceleration traces, but they organize into bands of transport strength across the map. In both cases the long-time evolution is diffusive, which indicates that Anderson localization is disrupted to different degrees. In the example of Figure~\ref{fig:fig1} the diffusivity in the diabatic case is about ten times larger than in the adiabatic case, which shows that the diabatic regime usually suppresses localization more strongly.

Accordingly, the acceleration trace provides a direct and robust diagnostic. Adiabatic motion produces sharp peaks in $a_{\rm cm}(t)$ that pass scale-invariant thresholds on relative height, prominence, width, and separation. Diabatic motion yields only broad undulations that fail these thresholds. The time of the first accepted peak defines the first switching time $\tau$. The density and prominence of accepted peaks correlate with the mean-squared acceleration ($\mathrm{MSA}$), which supplies a continuous calibration of adiabatic activity consistent with the kink classification.

\subsection{Phase Diagrams: Regime Map, Planckian Domain, and Quantified Coherence}

\begin{figure}[H]
  \centering
  \includegraphics[width=0.96\columnwidth]{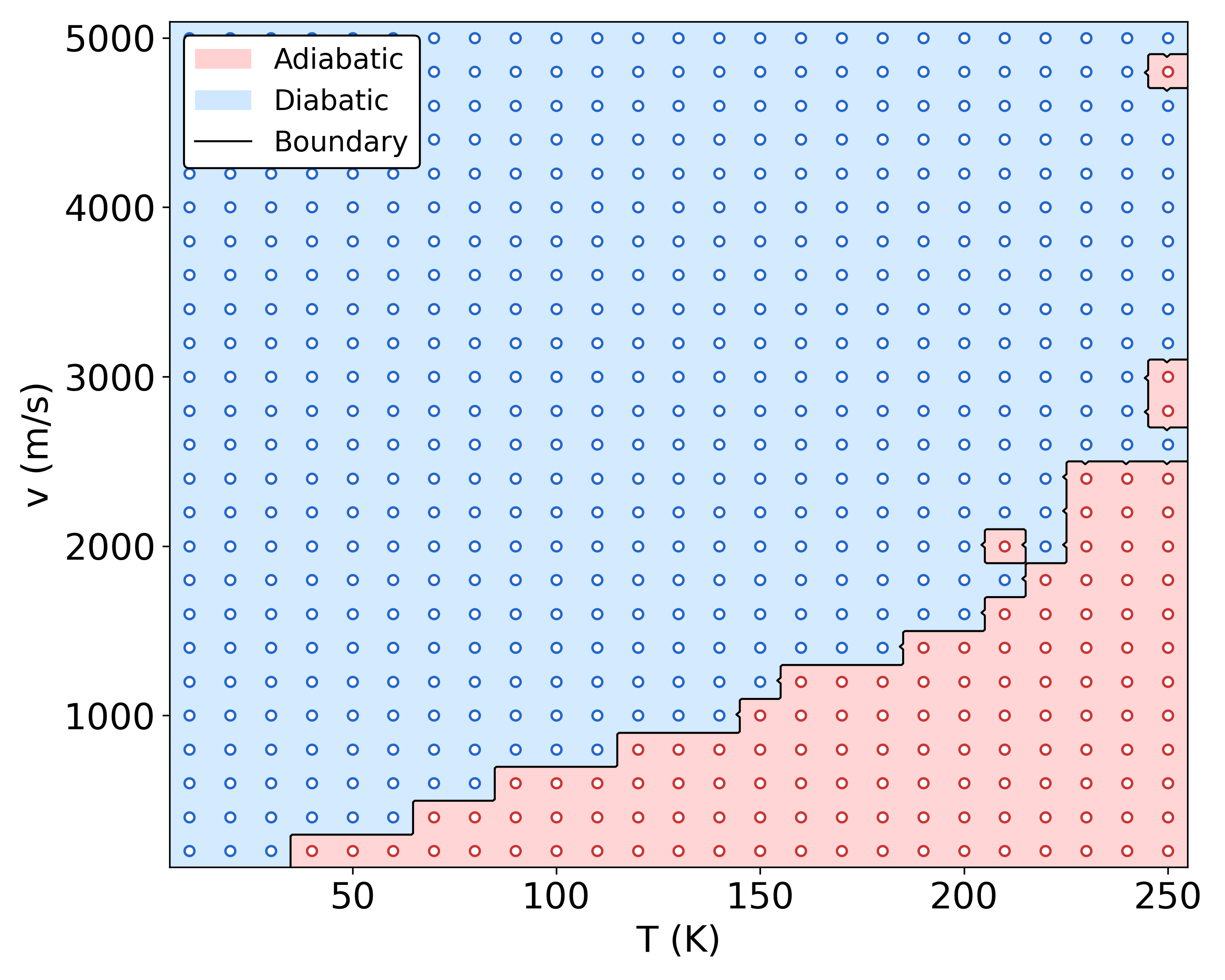}
  \caption{\label{fig:fig2}
  Adiabatic--diabatic regime map in the $(T,v)$ plane at fixed $g$ from the acceleration-based adiabatic criterion. Red symbols denote adiabatic points, blue symbols denote diabatic points, and the black staircase marks the boundary.}
\end{figure}

With the acceleration-based adiabatic criterion we map the $(T,v)$ plane at fixed coupling $g$ into two dynamical sectors. Figure~\ref{fig:fig2} shows the discrete adiabatic versus diabatic classification. As temperature increases or sound speed decreases the system shifts toward the adiabatic side, consistent with stronger temporal averaging of the deformation landscape \cite{https://doi.org/10.1002/lpor.202501144}.

\begin{figure}[H]
\centering
\includegraphics[width=0.96\columnwidth]{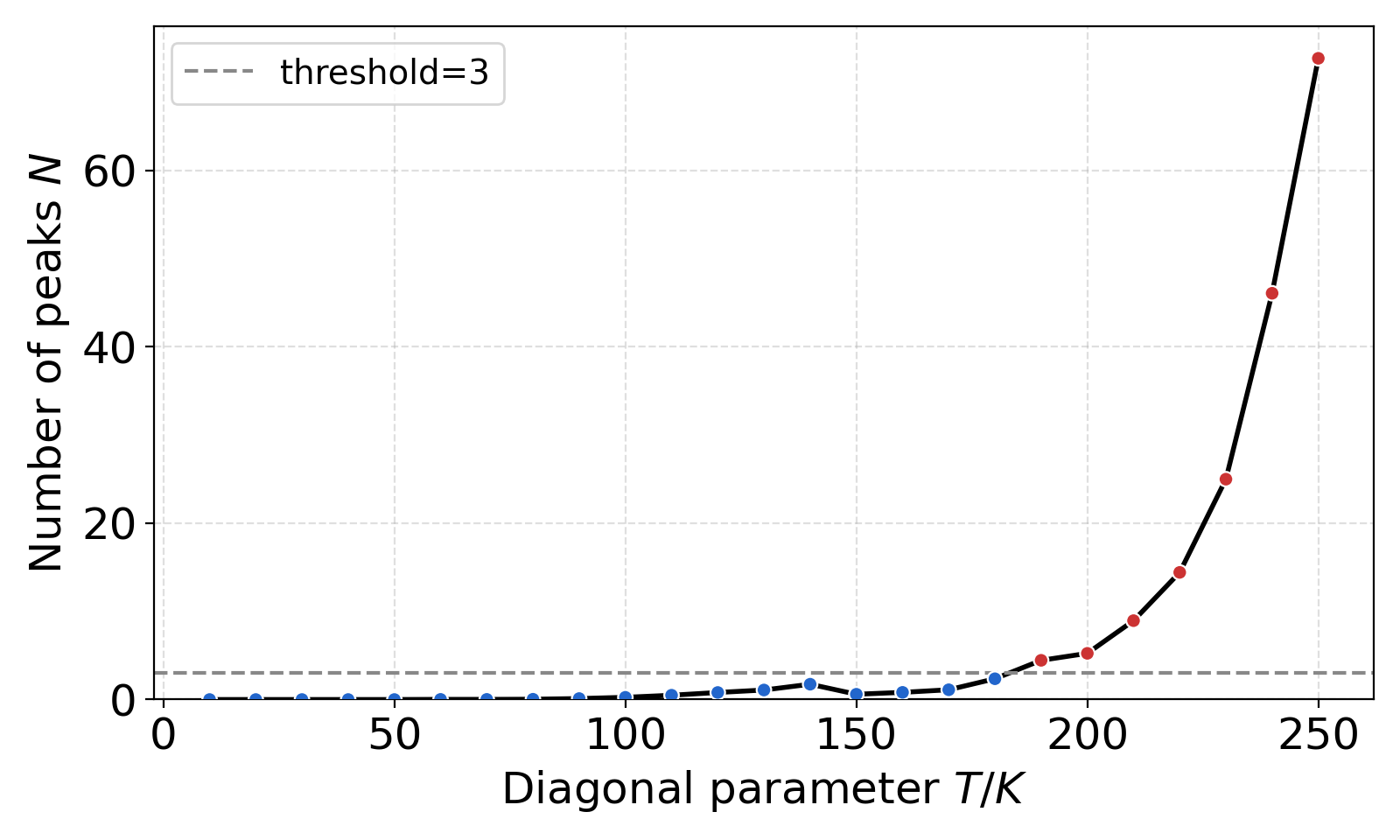}
\caption{\label{fig:fig3}
Average number of accepted peaks $N$ along the diagonal running from the top-left to the bottom-right of Figure~\ref{fig:fig2}. Each dot is the mean over 64 random thermal phase seeds, the gray dashed line marks the classification threshold $N=3$. Blue points lie in the diabatic sector, red points in the adiabatic sector. The sharp rise of $N$ shows a sudden onset of kink activity near the regime boundary.}
\end{figure}

To resolve more details at the boundary between the adiabatic and diabatic regimes, we examine the data points along the diagonal of Figure~\ref{fig:fig2} and count the accepted peaks over a fixed observation window, averaging over 64 random seeds (Figure~\ref{fig:fig3}). The curve exhibits a long diabatic plateau with $N\approx0$, then a very sharp rise where the classification changes, followed by rapid growth deep in the adiabatic side. Because the onset is sudden and the signal is large compared with baseline noise, moderate variations of the peak thresholds or the short smoothing window shift the crossing by only a few grid steps. This diagonal statistic shows that the boundary is controlled by the wave packet dynamics rather than by algorithmic tuning, and that the classifier is stable and physically meaningful.

\begin{figure}[H]
  \centering
  \includegraphics[width=0.96\columnwidth]{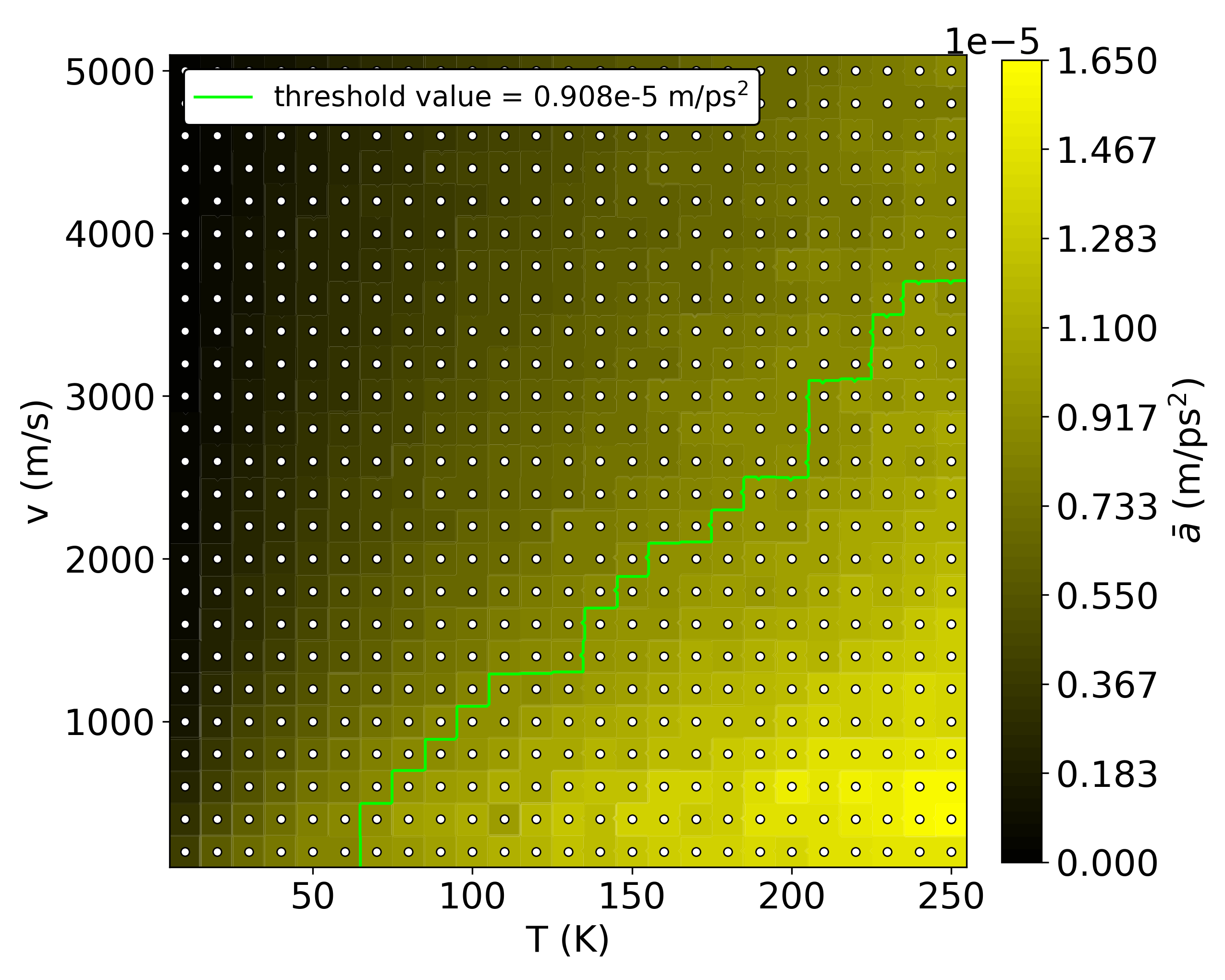}
  \caption{\label{fig:fig4}
  Mean-squared acceleration ($\mathrm{MSA}$) heat map on the $(T,v)$ grid. Color indicates $\mathrm{MSA}$ magnitude, with brighter yellow representing a more adiabatic system, and the green polyline shows a representative MSA contour.}
\end{figure}

We compare this discrete labeling with a continuous proxy based on the mean-squared acceleration. In Figure~\ref{fig:fig4}, a single $\mathrm{MSA}$ contour aligns with the regime boundary from the kink diagnostic, which calibrates adiabaticity on an absolute scale and validates the acceleration-based classifier.

\begin{figure}[H]
  \centering
  \includegraphics[width=0.96\columnwidth]{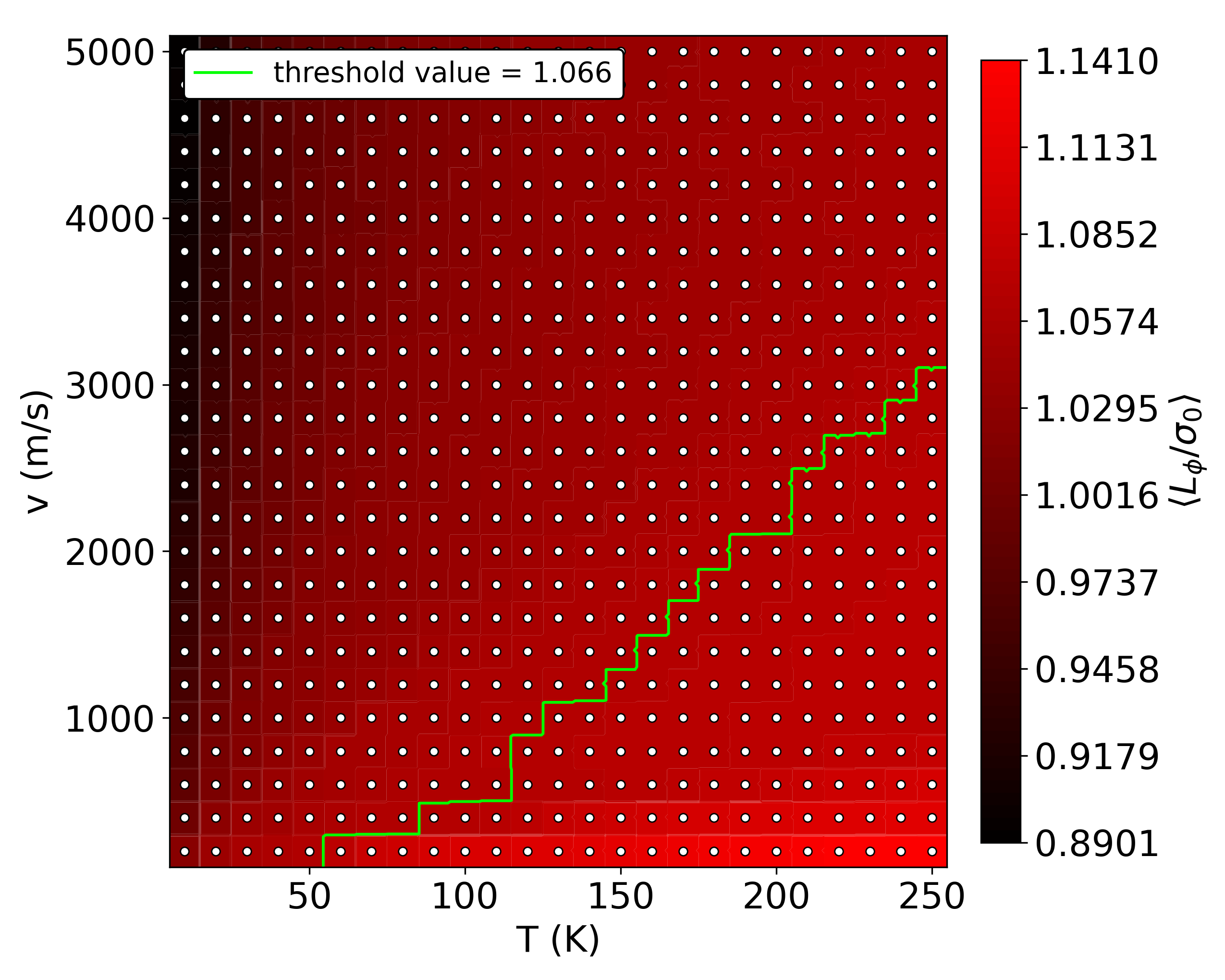}
  \caption{\label{fig:fig5}
  Coherence map reporting the dimensionless ratio $L_{\phi}/\sigma_{0}$ on the $(T,v)$ grid, with brighter red indicates a system with stronger coherence. Color encodes $L_{\phi}/\sigma_{0}$ and the green polyline marks a reference threshold.}
\end{figure}

We next relate dynamical regime, phase coherence, and Planckian diffusion. Anderson localization in one dimension relies on phase-coherent backscattering \cite{RevModPhys.57.287,PhysRevLett.55.2696}. Purely periodic motion of the landscape does not by itself erase coherence; dephasing requires fluctuations that carry effectively irreversible information \cite{10.1063/1.5054017}. In the quantum-acoustic setting the multi-mode deformation potential has thermal phases and a colored temporal spectrum, which act as dynamic disorder and promote dephasing \cite{PhysRevB.106.054311}, related crossovers in time-fluctuating random potentials have been observed in photorefractive media \cite{https://doi.org/10.1002/lpor.202501144}. Diabatic dynamics experience stronger exposure to these fluctuations and thus stronger dephasing, whereas adiabatic dynamics feature well-separated, resolvable channel switches with weaker dephasing. Figure~\ref{fig:fig5} quantifies the remaining coherence via the dimensionless ratio $L_{\phi}/\sigma_{0}$ and shows that the contour where $L_{\phi}/\sigma_{0}\!\sim\!1$ tracks the adiabatic--diabatic divide.

\begin{figure}[H]
  \centering
  \includegraphics[width=0.96\columnwidth]{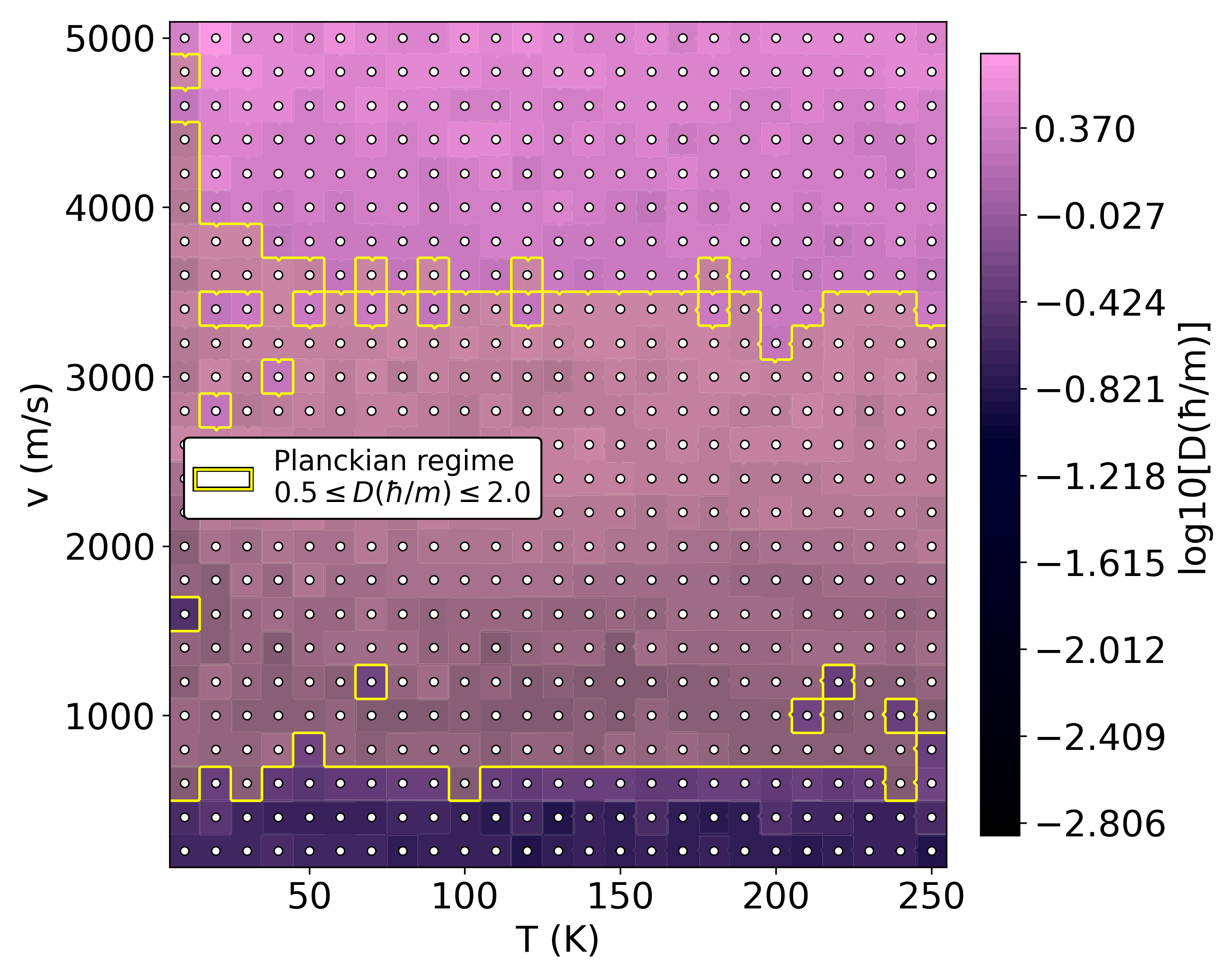}
  \caption{\label{fig:fig6}
  Diffusivity map showing $\log_{10}[D\,(\hbar/m)]$ on the $(T,v)$ grid, with brighter purple indicates a system with a larger diffusion coefficient. Yellow boxes highlight the Planckian domain where $0.5\le D\,(\hbar/m)\le 2.0$.}
\end{figure}

Finally, we connect coherence loss to transport. Figure~\ref{fig:fig6} delineates a broad Planckian domain where the dimensionless diffusivity $\alpha\!=\!Dm/\hbar$ is of order unity and varies only weakly with parameters \cite{RevModPhys.94.041002,doi:10.1126/science.1227612,Legros2019,Grissonnanche2021,zhang2024planckiandiffusionghostanderson}. This domain concentrates where Figure~\ref{fig:fig2} is diabatic and Figure~\ref{fig:fig4} indicates reduced $L_{\phi}/\sigma_{0}$. The three maps therefore tell a consistent story: dynamic, fluctuating deformation potentials suppress the interference corrections that produce localization and drive a crossover to diffusion with a nearly quantum-limited coefficient. Adiabaticity serves as an practical indicator of the residual coherence and predicts whether the dynamics remain localized or flow to Planckian diffusion.

\subsection{Insights into the $T$-Linear Resistance Effect in Strange Metals}

The Einstein relation connects diffusion to charge transport through
\begin{equation}
    \frac{1}{\tau_{\mathrm{tr}}}=\frac{k_{B}T}{mD}
\end{equation}
Inside the Planckian domain the diffusivity is nearly temperature independent and can be written as
\begin{equation}
    D=\alpha\,\frac{\hbar}{m},\qquad \alpha=\mathcal{O}(1)
\end{equation}
which immediately gives a $T$-linear relaxation rate
\begin{equation}
    \frac{1}{\tau_{\mathrm{tr}}}\simeq \frac{k_{B}}{\alpha\,\hbar}\,T
\end{equation}
Using the Drude relation, we obtain a $T$-linear resistivity
\begin{equation}
    \rho(T)=\frac{1}{\sigma(T)}\simeq
    \frac{m}{n e^{2}\tau_{\mathrm{tr}}}=
    \frac{m k_{B}}{\alpha\,n e^{2}\hbar}\;T
\end{equation}
Thus a broad $(T,v)$ region where $D$ is quasi-constant at the quantum scale yields a $T$-linear relaxation rate and a $T$-linear resistivity with a parameter-controlled prefactor \cite{albert1905a,PhysRevLett.107.066605,Hartnoll2015,RevModPhys.94.041002,doi:10.1126/science.1227612,Legros2019,Grissonnanche2021}.

\begin{figure}[H]
  \centering
  \includegraphics[width=0.96\columnwidth]{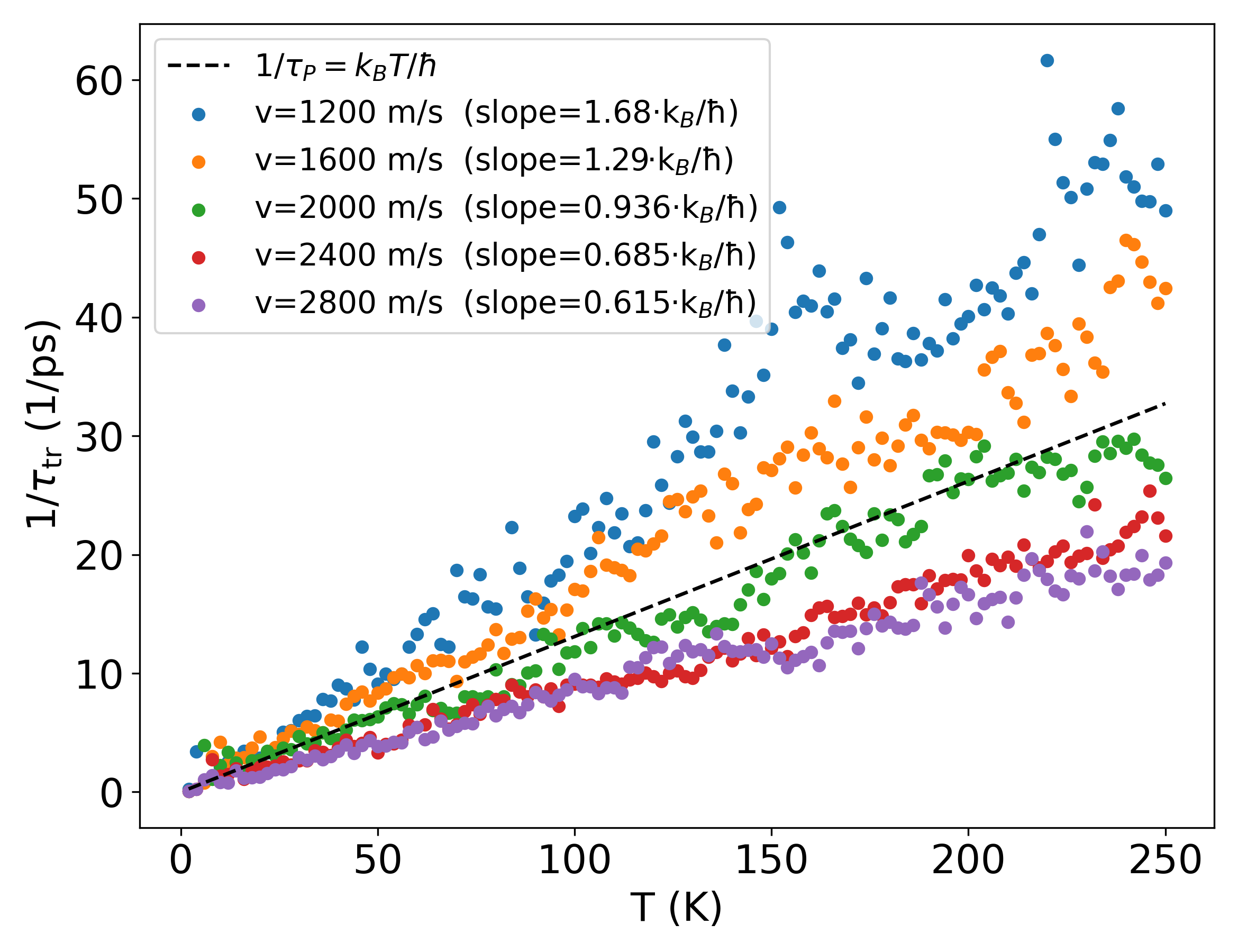}
  \caption{\label{fig:fig7}
  Temperature dependence of the transport relaxation rate $1/\tau_{\mathrm{tr}}$ at selected sound speeds $v$. A dashed reference shows the Planckian line $1/\tau_{P}=k_{B}T/\hbar$. Points follow an approximately linear trend whose slope decreases with increasing $v$. The lowest temperature point shown is 2 Kelvin.}
\end{figure}

Figure~\ref{fig:fig7} shows $1/\tau_{\mathrm{tr}}$ versus $T$ at several sound speeds $v$. Each series is close to linear at low temperatures, consistent with the Planckian plateau inferred for $D$. The dashed line marks the Planckian reference $1/\tau_{P}=k_{B}T/\hbar$ \cite{RevModPhys.94.041002}. The slope decreases with increasing $v$, indicating a larger effective $\alpha$ and therefore a reduced prefactor $k_{B}/(\alpha\hbar)$. These trends support a robust low-temperature $1/\tau_{\mathrm{tr}}\propto T$ tendency and provide a insight to $T$-linear resistivity in quasi-two-dimensional strange-metal settings \cite{Hartnoll2015,Legros2019,Grissonnanche2021}.

\section{Discussion}

The results presented above uncover a coherent picture that links adiabaticity, phase coherence, and quantum-limited diffusion. Within the quantum-acoustic framework a time-dependent deformation landscape with thermal modal phases drives the electronic wave packet \cite{doi:10.1073/pnas.2404853121,B_Kramer_1993,e26070552}. Our acceleration-based adiabatic criterion resolves two sharply distinct dynamical behaviors across the $(T,v)$ plane at fixed coupling $g$. In the adiabatic sector the wave packet performs a sequence of well separated channel switches that appear as kinks in spacetime maps and as sharp peaks in $a_{\rm cm}(t)$ [Figure~\ref{fig:fig1} (a) and Figure~\ref{fig:fig1} (b)]. In the diabatic sector the landscape varies rapidly and stochastically, the adiabatic condition fails, and the packet predominantly follows a diabatic branch without resolving individual avoided crossings, so the trajectory appears smooth and no accepted kinks are produced and peaks are suppressed [Figure~\ref{fig:fig1} (d) and Figure~\ref{fig:fig1} (e)]. The regime map in Figure~\ref{fig:fig2} and the continuous calibration by $\mathrm{MSA}$ in Figure~\ref{fig:fig4} show that the boundary is relatively sharp and tracks a single $\mathrm{MSA}$ contour. The phase-coherence map in Figure~\ref{fig:fig5} then demonstrates that adiabaticity is tightly correlated with the amount of phase coherence that the state retains, quantified by $L_\phi/\sigma_0$. Finally the diffusivity map in Figure~\ref{fig:fig6} reveals a wide Planckian domain where $\alpha\!=\!Dm/\hbar$ is of order unity and weakly dependent on parameters. This domain overlaps the diabatic sector and the low-coherence region, which closes the loop between dynamical regime, dephasing, and transport.

Conceptually, Anderson localization in one dimension arises from phase-coherent backscattering \cite{RevModPhys.57.287,PhysRevLett.55.2696}. A strictly periodic drive does not eliminate interference on its own. Dynamic disorder that carries fluctuations and randomness acts as an intrinsic dephasing environment \cite{10.1063/1.5054017}. Our simulations realize the latter through a multi-mode acoustic field with thermal phases \cite{PhysRevB.106.054311}. The adiabatic theorem and the Landau--Zener picture provide the operative mechanism \cite{Born1928,10.1098/rspa.1932.0165,Shevchenko2010_PhysRep_LZSReview}. When the sweep rate across local avoided crossings is small compared with the instantaneous gap squared, the system follows an instantaneous eigenstate and produces discrete channel switches. When the sweep is fast and carries stochastic fluctuations, the adiabatic condition is violated and the dynamics are diabatic. The state predominantly follows its diabatic branch without resolving individual avoided crossings, while random phase accumulation from the fluctuating potential dephases the interfering amplitudes and suppresses localization. The data in Figure~\ref{fig:fig2} through Figure~\ref{fig:fig6} show that increasing $v$ promotes nonadiabaticity by accelerating the temporal sweep of the landscape, whereas decreasing $T$ weakens fluctuations, pushes the dynamics toward the diabatic side, and reduces $L_\phi/\sigma_0$. The broad Planckian domain is then a natural consequence of dephasing that suppresses localization corrections and drives the system to a nearly quantum-limited diffusive fixed point \cite{RevModPhys.94.041002,heller2025quantumacousticsdemystifiesstrange}.

Figure~\ref{fig:fig8}is a cartoon of  the mechanism. We can simply accelerate the rate of change of the landscape to transition from he adiabatic regime into the diabatic regime. Figure~\ref{fig:fig8} (a) shows a local avoided crossing with red and blue diabatic levels. The adiabatic path follows the instantaneous eigenenergy and switches from the red channel to the blue channel, while the diabatic path remains on the original red channel. Figure~\ref{fig:fig8} (b) shows the common initial condition where the wave packet occupies the left pocket and is colored red. Figure~\ref{fig:fig8} (c) shows the passage stage. In the diabatic regime the state continues on the red dashed path and stays on the red channel. In the adiabatic regime the state would oscillate between the red and blue pockets, causing sudden hops, and sharp peaks appear in the center-of-mass acceleration. The intermediate adiabatic states are spendinf equal time in both wells and are colored purple purple. Figure~\ref{fig:fig8} (d) shows the outcome. The diabatic trajectory still resides on the red channel (left, higher), while the adiabatic trajectory has switched to the blue pocket, consistent with 

\onecolumngrid
\begin{center}
\setlength{\tabcolsep}{6pt}
\begin{tabular}{cccc}
\begin{overpic}[width=0.24\textwidth]{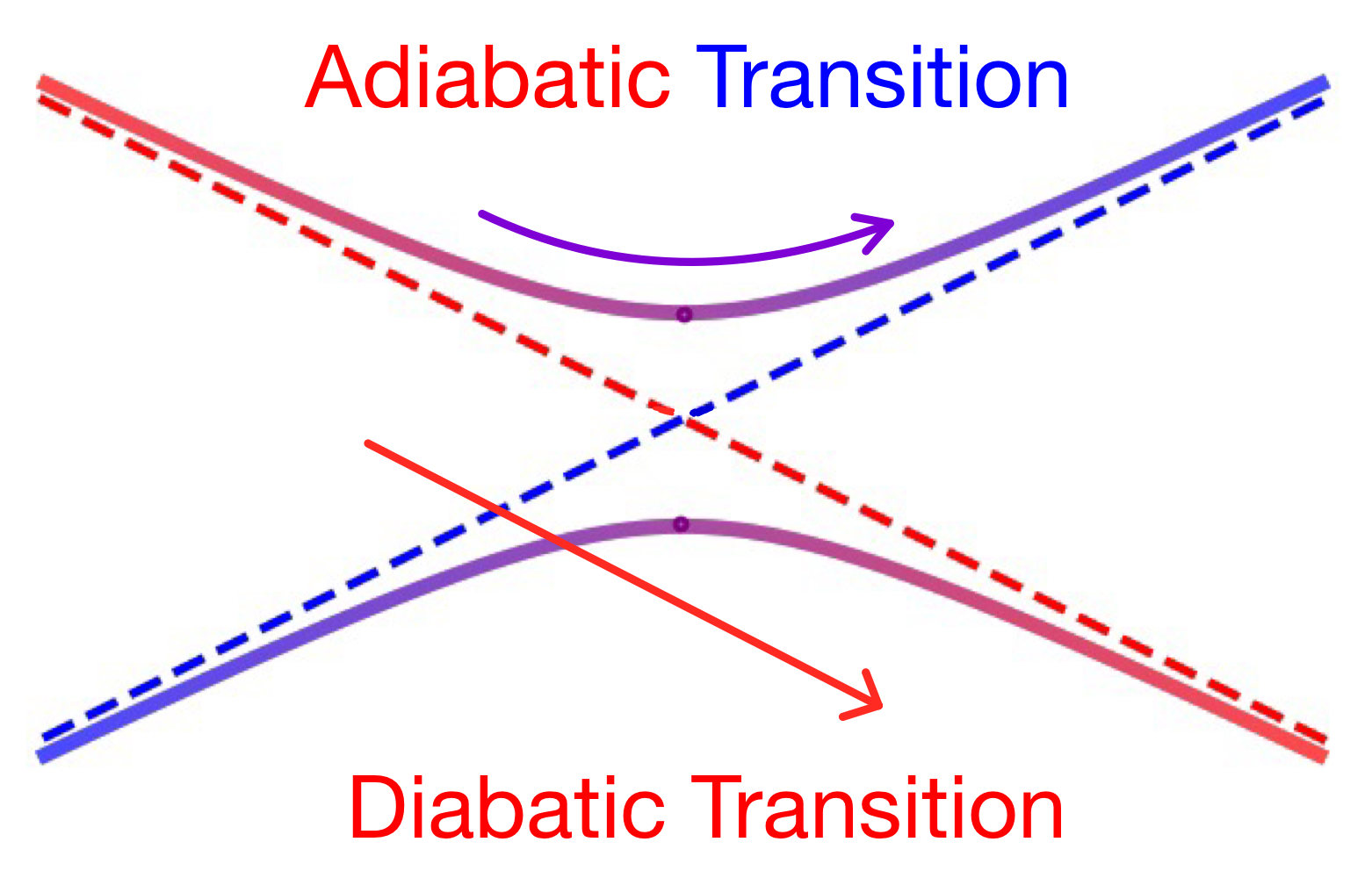}\put(0,78){\small (a)}\end{overpic} &
\begin{overpic}[width=0.24\textwidth]{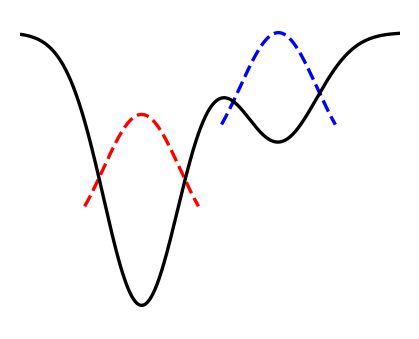}\put(0,78){\small (b)}\end{overpic} &
\begin{overpic}[width=0.24\textwidth]{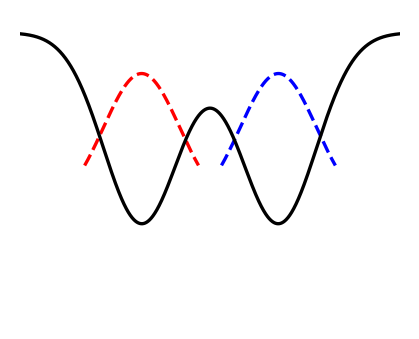}\put(0,78){\small (c)}\end{overpic} &
\begin{overpic}[width=0.24\textwidth]{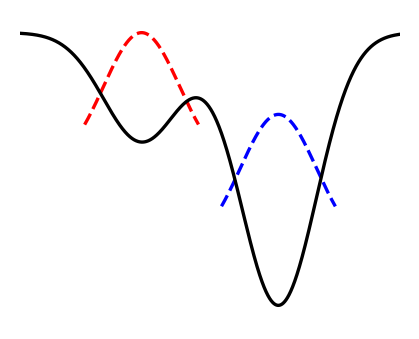}\put(0,78){\small (d)}\end{overpic}
\end{tabular}
\captionsetup{width=\textwidth, justification=RaggedRight, singlelinecheck=false}
\captionof{figure}{\label{fig:fig8}
Schematic of adiabatic and diabatic passage at a local avoided crossing and the associated kink generation.
(a) Avoided crossing with two diabatic levels shown in red and blue. The adiabatic path (purple arrow) follows the instantaneous eigenenergy and switches from the red channel to the blue channel, while the diabatic path (red arrow) stays on the original red channel.
(b) Initial stage: for both adiabatic and diabatic regimes, the wave packet occupies the red pocket.
(c) Passage stage: in the diabatic regime the state continues along the red dashed path and remains on the red channel, whereas in the adiabatic regime the state hybridizes between the red and blue pockets, an abrupt hop occurs, and a sharp peak appears in the center-of-mass acceleration, so the intermediate state appears purple.
(d) Final stage: the diabatic trajectory still resides on the red channel, while the adiabatic trajectory has switched to the blue pocket.}
\end{center}
\twocolumngrid

\noindent
kink generation and peak activity in the acceleration trace. This diabatic regime correlates with small $L_\phi/\sigma_0$ in Figure~\ref{fig:fig5} and with the Planckian domain in Figure~\ref{fig:fig6}. The wide extent of that domain indicates that the quantum-limited diffusive response is not a fine tuned feature but rather an emergent outcome of fluctuating dynamic disorder \cite{zhang2024planckiandiffusionghostanderson}.

In summary, adiabaticity provides an practical indicator of phase coherence for an electron driven by a quantum-acoustic landscape. Our acceleration-based adiabatic criterion cleanly resolves adiabatic and diabatic sectors across the $(T,v)$ plane. A broad Planckian domain emerges where the dimensionless diffusivity $\alpha\!=\!Dm/\hbar$ is of order unity and only weakly depends on the parameters. The regime, coherence ($L_\phi/\sigma_0$), and transport maps are mutually consistent, supporting a dephasing-controlled crossover from Anderson localization to Planckian diffusion. These findings motivate experiments that correlate phase-coherence measures with transport in low-dimensional materials subject to engineered dynamic disorder and predict a sound-speed dependence of the slope of $1/\tau_{\rm tr}$ versus $T$ testable via strain or alloying \cite{Schwartz2007,Billy2008,Roati2008,doi:10.1126/science.1209019,Jendrzejewski2012,White2020,Hu2008}. Extensions to resolve the microscopic mechanism by which the electron-lattice dynamics self organize into a polaron-like deformation potential well, and then leverage this mechanism to investigate interacting carriers are natural next steps.

\section{Acknowledgments}

This work was supported by the U.S. Department of Energy under Grant No. DE-SC0025489. A.A. acknowledges financial support from the Sabanci University President's Research Grant with project code F.A.CF.2402932. A.M.G. thanks the Studienstiftung des Deutschen Volkes for financial support. J.K.-R. thanks the Oskar Huttunen Foundation for the financial support.

\bibliographystyle{unsrt}
\bibliography{refs}

\end{document}